\newif\ifTR  
\TRtrue

\ifTR
\documentclass[conference,12pt,onecolumn]{IEEEtran}  
\IEEEoverridecommandlockouts           
\renewcommand{\baselinestretch}{1.15}
\else
\documentclass[10pt,sigconf,letterpaper]{acmart}

\AtBeginDocument{%
  \providecommand\BibTeX{{%
    \normalfont B\kern-0.5em{\scshape i\kern-0.25em b}\kern-0.8em\TeX}}}

\setcopyright{acmcopyright}
\copyrightyear{2019}
\acmYear{2019}
\acmDOI{10.1145/3366615.3368351}

\acmConference[ACM WOC'19]{ACM WOC'19}{December 09--13, 2019}{Davis, CA, USA}
\acmBooktitle{ACM WOC'19, December 09--13, 2019, Davis, CA, USA}
\acmPrice{15.00}
\acmISBN{978-1-4503-7033-2/19/12}
\fi

\usepackage{graphicx,comment,amsfonts,amsmath}
\usepackage{url}

\def\IEEEproof{proof}
\def\E{{\sf E}}

\def\P{{\sf P}}

\def\ud{\underline{d}}

\def\Jcal{\mathcal{J}}
\def\Mcal{\mathcal{M}}
\def\Ncal{\mathcal{N}}
\def\Ical{\mathcal{I}}
\def\Rcal{\mathcal{R}}

\def\eg{{\em e.g.},~}
\def\ie{{\em i.e.},~}

\newtheorem{theorem}{Theorem}[section]
\newtheorem{corollary}{Corollary}[section]
\newcommand{\beqa}{\begin{eqnarray*}}
\newcommand{\eeqa}{\end{eqnarray*}}
\newcommand{\be}{\begin{eqnarray}}
\newcommand{\ee}{\end{eqnarray}}

\ifTR
\newcommand{\qed}{\raisebox{.65ex}{\fbox{\rule{0mm}{0mm}}}}
\fi

\begin{document}

\title{Overbooking Microservices in the Cloud}

\ifTR
\author{George Kesidis\\
Pennsylvania State University
University Park, PA, 16802\\
gik2@psu.edu
}
\else
\author{George Kesidis} 
\affiliation{
\institution{Pennsylvania State University}
\city{University Park}
\state{PA} 
\postcode{16802}
}
\email{gik2@psu.edu}
\fi

\ifTR
\maketitle
\fi
\begin{abstract}
We consider the problem of scheduling 
serverless-computing instances such as Amazon Lambda functions,
or scheduling microservices within (privately held) virtual machines (VMs). Instead of a quota per tenant/customer, we assume demand for Lambda functions is modulated by token-bucket mechanisms per tenant. Such quotas are due to, \eg limited resources (as in a fog/edge-cloud context) or to prevent excessive unauthorized invocation of numerous instances by malware. Based on an upper bound on the stationary number of active ``Lambda servers" considering the execution-time distribution of Lambda functions, we describe an approach that the cloud could use to overbook Lambda functions for improved utilization of IT resources. An earlier bound for a single service tier is extended to multiple service tiers. 
For the context of scheduling microservices in a private setting,
the framework could be used to determine the required VM resources for a 
token-bucket constrained workload stream.
Finally, we note that the looser
Markov inequality may be useful in settings where
the job service times are dependent.
\end{abstract}

\section{Introduction}

Public-cloud computing 
is conducted through 
Service-Level Agreements (SLAs), including pricing policies.
Also,
there is limited information-sharing regarding workloads
between tenants\footnote{a.k.a. customers or users} and the operator/provider
of a neutral public cloud \cite{HotCloud16-neutrality}.
Though public-cloud operators  may seek to maximize their revenue  and 
minimize their operating (including amortized capital) expenditures, 
they may be forced to 
treat tenants ``fairly" according to future neutrality regulations.
Moreover, it may not be permitted to profile individual tenants, though
it may be permitted to profile, \eg a particular service spanning
all tenants that use it.

A variety of cloud-computing services have been broadly classified as
Infrastructure-as-a-Service (IaaS) such as
Virtual Machines (VMs), Platform-as-a-Service (PaaS) including
Function-as-a-Service (FaaS) such as Amazon Lambda
``serverless" computing, and Software-as-a-Service (SaaS) such as 
GCE's TensorFlow.   
We focus herein on PaaS as offered by 
AWS (Lambda), GCE, Azure and IBM Cloud. In the following, we will
call PaaS invocations Lambda functions or Lambda service instances.

Rather than renting reserved resources through a VM, under
serverless computing multiple stateless Lambda functions
are submitted  by a tenant for
execution in a provisioned container.
AWS Lambda service tiers are based on 128MB units of memory, with
2 vCPU allocated per 3GB memory ($\frac{1}{12}$ vCPU per memory unit).
Cost per tier is based on units of memory times the time that  the
Lambda invocation is active. 
State spanning
plural Lambda invocations is externalized, \eg managed by a ``master" 
or ``driver" VM or stored in AWS S3 or Single Queue Service (SQS);
also see \cite{splitserve19}.


Lambda service instances  typically require on the order
of tens of milliseconds to a few minutes execution time
\cite{Swift18,Stein18,RSBarga};  in the lower range of
execution times, cold-start spin-up overhead (including data acquisition) 
can be substantial.
But to avoid such delays, a (cloud controlled) container may
persist after a Lambda function finishes execution 
in anticipation of additional demand by the same 
tenant \cite{Swift18}. 
However, reserving the IT resources of dormant/paused containers for future
invocations by the same tenant
could be very resource inefficient.

In the following, we assume that an idle
IT resource bundle for Lambda service,
considered to be a ``Lambda server", can be
used by any tenant at any time as permitted by their SLA.
A disadvantage
is that there may not be sufficient isolation among different cloud
tenants under this assumption \cite{AWS-Lambda-isolation}, 
\eg presently, important data may be leaked from
one tenant (whose Lambda function terminates) to another (whose Lambda
function shortly thereafter commences in the same cloud-managed VM)
through memory side-channels  (\ie the memory used by a Lambda function is
not erased, or an equivalent operation performed, upon its termination).

Some providers limit the number of simultaneous cloud-function
service-instances per tenant, \eg
AWS {\em concurrency} limits
are described in \cite{AWS-concurrency}.
There are security and cost risks to the
tenant associated with autoscaling
due to faults, the actions of intrusive
malware, deliberate Denial-of-Service (DoS) attacks,
or due to nominal but unexpected resource congestion (flash crowds)
\cite{AWSPrimeDay18}.
Concurrency limits may control  such risks\footnote{A large tenant
with several concurrent applications could similarly employ a 
token-bucket mechanisms to control how an individual
applications launches Lambda functions.}.

\ifTR
A tenant may explore performance/cost tradeoffs (including
security) for serverless computing by 
implementing the same function differently, \eg to reduce memory use. 
We also note that burstable/bursting  and spot/preemptible VMs 
are less expensive
to rent, the former having
only intermittently available CPU (and network I/O)
resources as governed by a token-bucket mechanism \cite{SIGMETRICS17}.
\fi

Also,
the cloud provider generally wishes to operate their 
infrastructure efficiently.
Efficient cloud operation, and associated potential cost savings for tenants, 
will be particularly
important in edge/fog  computing settings where: prices are
generally much higher, concurrency limits per tenant are likely to be
stricter, and servers mounting Lambda functions
are likely to be shared among different tenants
(rather than dedicated to individual tenants).

This paper focuses on the problem of cloud-side scheduling
and consolidation of Lambda service instances, particularly
principled approaches 
to overbook resources so as to improve  utilization efficiency
and thus maintain greatest possible 
 service availability to 
tenant customers. So, from the tenant's point of view, 
the edge-cloud Lambda service will be more
dependable, particularly for autoscaling,
 notwithstanding congested (and costly) IT 
resources. 

The following framework may also be useful in a more ``private"
setting-up of tenant-rented Virtual Machine (VM) resources housing
containers executing a microservice-workload stream.
Here, the aim could be to determine the number of VMs and
their sizes so as to limit the amount of autoscaling while using
these procured resources efficiently.

Finally, note that individual job service times may vary greatly,
even in a microservice setting. So, limiting the job arrival
stream by a token-bucket mechanism and just bounding the job execution
times can lead to very inefficient use of resources.
This motivates a simple statistical model for job service times.
(Note that in packet switching, packet sizes are known {\em a priori}.)

This paper is organized as follows.
Related work is discussed in Section \ref{sec:related}.
The problem is set up in Section \ref{sec:back} and 
a no-blocking condition is given when 
a service quota is replaced by a
token-bucket  mechanism governing concurrency,
\ie governing how tenants may request
homogeneous Lambda service instances.  
In Section \ref{sec:loss}, we show
how admission control
can be relaxed considering empirical Lambda-function execution times resulting
in more efficient use of resources.
An extension to multiple service tiers based in part on 
allocated resources per Lambda service instance is discussed in 
Section \ref{sec:multiclass}.
The paper concludes with a 
discussion of 
future work in Section \ref{sec:fw}.

\section{Related Work}\label{sec:related}

For decades, token-bucket mechanisms
have been used to control the resource utilization
of a workload stream. In a packet-switching  context,
\eg \cite{Cruz95,2R3CM,1R3CM}, the tasks (packet-header processing
and packet transmission) have very predictable sizes\footnote{IP packet
lengths are simply given in their headers.} compared  
to workloads of a general-purpose
CPU, call center, {\em etc.}
For scheduling purposes in the latter cases, 
token-bucket controls at the task level may be augmented
by statistical models profiling task execution times, \eg
\cite{KesTas00,Pacheco-SanchezCSMPD11,Delimitrou13,Delimitrou16}. 
In some cases, predictable workloads can be overbooked to
improve resource-utilization efficiency.
Some prior work on resource  overbooking has been
based on chance constraints, \eg
involving second-order statistics
\cite{HotCloud16-neutrality,Cohen17}.

Though we assume herein that Lambda-function invocations are limited
by a deterministic token-buck mechanism, resource allocation
to Lambda functions will also depend on the distribution of
their execution times, as estimated by the cloud. 
Such estimates could be continually updated over time,
as new Lambda-function execution-time statistics are collected.
For example, a classical maximum-likelihood approach can be
used to fit a sliding time-window of the most recent 
cloud-function execution times to a parameterized distribution model,
\eg of the Gamma \cite{Minka02} or Weibull type.
In an online setting,
if updates are based on observation batches,
the old approximate service-time distribution, $\hat{p}$,
and the one based on the most recent batch of 
observed Lambda execution/service times,
$\hat{q}$, could be combined in a simple first-order autoregressive manner,
$\alpha\hat{p}+(1-\alpha)\hat{q}$,
where forgetting factor $\alpha$ is such that $0<\alpha<1$.

\section{Problem set-up and a no-blocking condition}\label{sec:back}

Consider available
 resources of a  set $\Ical$ of heterogeneous physical servers,
including 
resources unused by existing IaaS instances (VMs)\footnote{Considering
the fleeting nature of Lambda service, some cloud operators may be tempted to 
use idling capacity reserved for IaaS for Lambda service.}.
Let $c_{i,r}$ be the amount of IT resource of type $r\in \Rcal$ (\eg 
$\Rcal=\{$vCPUs, memory, network I/O$\}$) available for 
Lambda service on server $i\in \Ical$.
In the following, 
$\min_r$ will be short for $\min_{r\in \Rcal}$, 
$\sum_i$ will be short for $\sum_{i\in \Ical}$, {\em etc.}

Consider a set $\Ncal$ of tenant-customers of a common type of Lambda service,
with 
$d_r$ being the amount of
type $r\in \Rcal$ resource allocated per invocation as prescribed
by the SLA.
In the following, we assume tenant SLAs stipulate
\begin{itemize}
\item IT resources allocated per invocation of the common
type of Lambda service, $\{d_r\}_{r\in\Rcal}$,
\item a maximum execution/activity time $S_{\max}$ per invocation,
\item and some limit to the rate at which tenants can request
different Lambda service instances.
\end{itemize}

For the case of tenant demand for a single
type of Lambda service,
we can consider each 
available $|\Rcal|$-vector of resources $\ud$ from the
physical server pool $\Ical$ as a ``Lambda server"  that
pulls in work when idle, \eg \cite{MB17}.  

Suppose that 
there are $K$ such servers available:
\be\label{K-def}
K & = & \sum_{i\in\Ical} \min_{r\in \Rcal} \left\lfloor 
\frac{c_{i,r}}{d_r}\right\rfloor.
\ee
Generally, $K$ is time-varying
but at a longer time-scale than that of individual Lambda-service 
lifetimes or of the time between successive Lambda-service invocations.

\subsection{A quota system}

First
note that if there is a simple quota, $K_n<K$, 
on the number of active Lambda invocations
for tenant $n$, then by Little's formula, $K_n/\E S_n$ is an upper bound
on the {\em mean} rate at which that tenant can request service, where
the random variable $S_n$ is distributed as the execution time of tenant $n$'s
Lambda functions.
Furthermore, if tenant $n$'s service-request process 
is modeled as Poisson, then the Erlang blocking formula applies
\cite{Wolff89}.  

In the following, we do not assume a Poisson model for
service request processes.

The cloud may overbook resources by, \eg online estimating the mean and variance
of the total number of active Lambda servers $Q\leq K$, respectively
$\widehat{\E Q}$ and $\widehat{\mbox{var}(Q)}$, using, \eg a simple
autoregressive mechanism. Admission control could be 
based on the current 99\%-confident estimate 
$K - \widehat{\E Q}-3\sqrt{\widehat{\mbox{var}(Q)}}$ of available Lambda servers.
SLAs should capture how such overbooking approaches may 
sometimes result in blocking 
of within-quota requests for Lambda service.

\subsection{Demand constrained by token bucket regulators}

Instead of a simple quota on the number of active invocations per tenant,
the cloud can accommodate batch Lambda-service requests while 
effecting control on such a system by applying 
token-bucket  allocators. For example, a dual token-bucket
allocator permits  only
\be\label{dtb}
g(t) & = & \min\{b+\pi t,~\sigma+\rho t \}
\ee
requests for Lambda service over any time-interval of length $t$,
with peak rate larger than sustainable rate,
$\pi>\rho>0$, and the maximum burst size at the sustainable
rate greater than the number of  simultaneous new Lambda-service requests
that can be submitted, $\sigma>b$.

Note that, just as in a fixed quota system, 
every tenant
is immediately aware of how many new Lambda service instances they can
invoke at any given time
based on their current token-bucket state.

Different tenants may engage in different service tiers
$\Jcal$ corresponding
to different dual-token bucket mechanisms governing their rate of
Lambda-service requests.  Let $j(n)\in\Jcal$ be the service tier of
active tenant $n$ corresponding to burstiness curve
$g_{j(n)}$. 
The maximum service time per Lambda-service
invocation, $S_{j(n),\max}$, is also assumed to be stipulated in SLAs.

The following constraint
\be\label{no-blocking}
\sum_n g_{j(n)}(S_{j(n),\max}) & \leq & K 
\ee
will imply that all Lambda-service requests satisfying (\ref{dtb})
will be invoked upon request \cite{Cruz95}. So,
if the number of available servers is $K$, and $\mathcal{N}$ is the
current set of active tenants, then a new tenant at service tier $j\in \Jcal$
is admitted only if
\beqa
g_j(S_{j,\max}) & \leq & 
K - \sum_{n\in\mathcal{N}} g_{j(n)}(S_{j(n),\max}).
\eeqa

If the tiers are designed so that there is an ``atomic" tier
$1\in \Jcal$ based on its burstiness curve $g_1$
(\ie for every tier $j\in \Jcal$, $j$ is an integer such that
$g_j = j g_1$),
then a price $p_j$ for tier-$j$ invocations satisfying
$p_j < j p_1$ would correspond to a 
volume discount.

\section{Overbooking  based 
on service-time distribution for a single service tier}\label{sec:loss}

Consider a single service tier.
As (\ref{no-blocking}) may be very conservative,
the cloud may instead profile the service-time distribution $S$ across
all tenants and service tiers and employ  our
\ifTR
Theorem 2 of \cite{KesTas00} (reinterpreted in Appendix A).
\else
Theorem 2 of \cite{KesTas00} (reinterpreted in Appendix A of \cite{lambda-netcalc-arxiv}).
\fi
For an infinite server system,
this result uses the Chernoff bound to show 
that the probability that
the number of busy servers $Q$ exceeds $K$,
\be
\lefteqn{\P(Q>K) ~\leq ~ \Omega(\mathcal{N}) ~:=~
 \exp\Bigg(-\sup_{\theta>0}\Bigg\{\theta (K-g(0))
} & &  \nonumber \\
& & ~~~ -\int_{g(0)}^{g(S_{\max})} \log(\Phi(x)\mbox{e}^\theta + 1-\Phi(x))
\mbox{d}x \Bigg\}\Bigg), 
 \label{yes-blocking}
\ee 
where $\Phi(x) := \P(g(S)>x)$ and $P(S=0)=0$ is assumed.

Note that the looser Markov inequality
of Corollary \ref{cor:Markov} in Appendix A
(relying only on common mean service times)
does not require independent service times.

In the following, 
this theorem is extended to multiple service tiers.

If the distribution 
of $S$ based  on recent Lambda-service invocations
is continually estimated,
then $\Phi$ and, in turn, 
the bound $\Omega(\mathcal{N})$
can be numerically computed for the given set of active tenants $\Ncal$.
If there is  a small tolerable aggregate blocking probability of $\varepsilon>0$
(a quantity that could be stipulated in SLAs), 
a new tenant $n'$ is admitted if 
\beqa
\Omega(\mathcal{N}\cup\{n'\})  & \leq & \varepsilon,
\eeqa
here assuming that the new tenant $n'$ will have negligible impact
on the (collective) execution-time distribution.

Again, Lambda service instances  typically require on the order
of tens of milliseconds to a few minutes execution time
\cite{Swift18,Stein18,RSBarga}\footnote{Note that the
hour-scale ``lifetimes" of Fig. 9 of \cite{Swift18}  are the
{\em overall} lifetimes of the lambda functions, spanning
plural such execution (service instance) times separated by 
dormant/pause periods.}.
For a numerical example,
suppose the cloud models  
Lambda-service instances as having
independent execution-times (lifetimes) $S$ 
that are (bell-shaped and  non-negative) 
Weibull distributed  with scale parameter 
1 and shape parameter 5 so that the mean is 0.915 (minutes), and 
truncated so that $S_{\max}=1.4$ (at which point this 
Weibull density is approximately zero).
Also suppose the collective burstiness curve is $g(t)=5+100t$, \ie just
a single token-bucket mechanism. Thus, the {\em mean} rate of
{\em invoked} requests is less than 100 per second.

\begin{figure}
\centering
\includegraphics[width=3.25in]{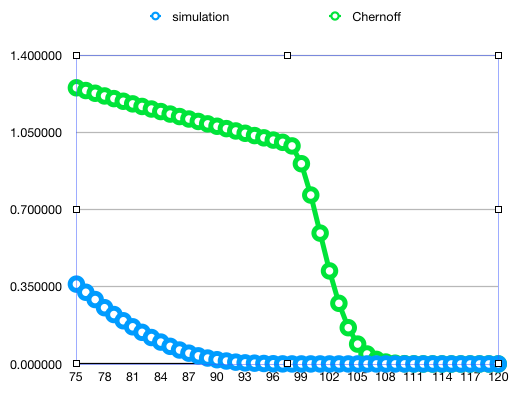}
\caption{$\P(Q>K)$ evaluated by simulation 
and its Chernoff bound (\ref{yes-blocking})
versus $K$ for Poisson batch requests. Here $\E Q=71$.
}\label{fig:poisson}
\end{figure}

\begin{figure}
\centering
\includegraphics[width=3.25in]{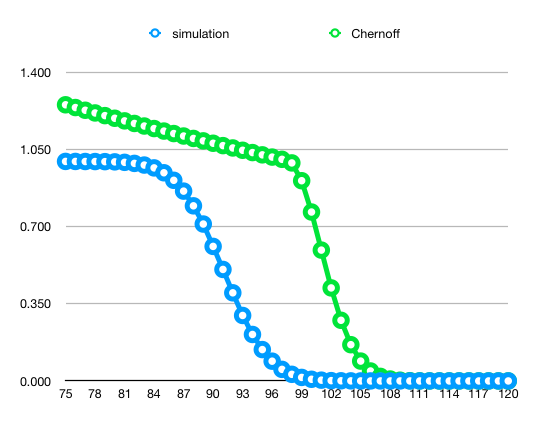}
\caption{$\P(Q>K)$ evaluated by simulation 
and its Chernoff bound (\ref{yes-blocking})
versus $K$ for deterministic batch requests. Here $\E Q=92$.
}\label{fig:deterministic}
\end{figure}

In one numerical example,
we took two cases for demand. The first was a maximal $g$-permitted
deterministic demand process wherein a batch of $5$ instance requests 
were made every $\frac{1}{20}$ second.
In the second case, we simulated a Poisson process with mean rate
20 and batches of 5 instances were requested for each Poisson
arrival. In the Poisson
case, some requests did not satisfy the burstiness curve $g$ and were
dropped so that
the average admitted batch size was only $3.9$ (so, a mean rate of
$20\times 3.9=78$  invoked requests per second).
The mean number of occupied servers by simulation (or Little's
formula), $\E Q= 92$ for deterministic batch requests and
$\E Q=71$ for Poisson batch requests.
Numerical results are given in Table \ref{table:netcalc-loss} 
and Figures \ref{fig:poisson} and \ref{fig:deterministic}. 
We see that the Chernoff bound (\ref{yes-blocking}) 
does reasonably well indicating
the number of required servers when the burstiness curves well
reflect demand and blocking tolerance $\varepsilon$ is small,
while (\ref{no-blocking}) is very conservative
even in this case.

We numerically found that the Markov inequality given in
\ifTR
\cite{KesTas00} (also see Appendix A),
\else
\cite{KesTas00},
\fi
though much
easier to compute than (\ref{yes-blocking}),
is much more conservative even than (\ref{no-blocking}).
This said, it is relevant to cases where the service times $S_i$ are
dependent (and $\E g(S)< K$, of course).

\begin{table}
\centering
\begin{tabular}{c|c|c|c}
min $K$ s.t. & &  &  \\ 
$\P(Q>K)<.01=\varepsilon$  & simulated  & (\ref{yes-blocking}) & (\ref{no-blocking}) \\ \hline
deterministic  & 100 & 108 & 145 \\  \hline
Poisson        & 92  & 108 & 145 
\end{tabular}
\caption{The minimum number of servers $K$ required so that 
$\P(Q>K)<0.01=\varepsilon$,
\ie the stationary probability that
the number of occupied servers $Q>K$ is less than one percent, for
simulated system and according to the Chernoff bound (\ref{yes-blocking})
and the no-blocking bound (\ref{no-blocking}).}\label{table:netcalc-loss}
\end{table}


\subsection{Overbooking based on empirical weak burstiness curves on 
service-request process}

The bound on blocking probability $\Omega(\mathcal{N})$ may still be conservative
considering that many tenants may not request at close to the maximum rates given
by (\ref{dtb}) of their service tiers. To this end, an {\em empirical} 
service-request
envelope $\hat{g}$ can be estimated for all currently active tenants, 
and $g$ can be replaced by $\hat{g}$ in the definition of $\Phi$.
(That is, $\hat{g}$ can inform the burstiness curve requested by the tenant
of the cloud.)
Here, $\hat{g}$ can be any increasing, concave and nonnegative function.

To this end, consider the notion of a ``weak" burstiness curve  constraint 
involving a small positive confidence parameter $\delta<1$
\cite{Low91}.
Given the aggregate number of service requests over time interval $(s,t]$,
$A(s,t]$, one can track {\em virtual queues}  (\eg \cite{1R3CM,2R3CM})
$V_r(t) = \max_{s\leq t} A(s,t]-r(t-s)$ for different 
service rates $r\leq \pi$. 
For each virtual queue, we can estimate 
minimal $\hat{\sigma}_r$ such that
\beqa
\P(V_r > \hat{\sigma}_r) & < & \delta.
\eeqa
In particular, the maximum simultaneous aggregate request observed
$\hat{b}=\hat{\sigma}_r$ for $r=\pi$.
Note that $\hat{\sigma}_r \leq \hat{\sigma}_{r'}$ if $r > r'$.
Thus, we can approximate (concave)
\beqa
\hat{g}(t) & = & \min_r \hat{\sigma}_r + tr .
\eeqa

\section{Discussion:
Multiple service tiers for one resource pool}\label{sec:multiclass}

Consider the case where the aggregate demand 
of tier $j\in\mathcal{J}$ 
has service-request burstiness curve $g_j$
and i.i.d.  execution-times $\sim S^{(j)} \leq S^{(j)}_{\max}$ such that
$\P(S^{(j)}=0)=0$ $\forall j\in\Jcal$.
 Assume arrival and service
processes of each tier are mutually independent.
Furthermore,
suppose each Lambda service instance of type $j\in\Jcal$ requires
an amount $d_{j,r}$ of resource of type $r\in\Rcal$.
For all $j\in\Jcal$ and $r \in \Rcal$, let
\beqa
q_{j,r} & = & Q_jd_{j,r} 
\eeqa
be the total stationary amount of resource of type $r$ 
allocated to active Lambda service instances of type $j$ for
an infinite resource system.

Assume that resources for Lambda service are from a single pool
(physical server), $i$. 
Recall that  the amount of type-$r$ resource available is $c_{i,r}$.

\begin{corollary}\label{cor:mc}
	For physical server $i$,
\beqa
	\lefteqn{\P\left(\max_{r\in\Rcal} \frac{ \sum_j q_{j,r}}{c_{i,r}} > 1\right)} & & \\
 & \leq
& \exp\left(-\sup_{\theta>0}\left\{\theta - \sum_{j\in\Jcal}
\Mcal_j\left(\theta\max_{r\in \Rcal}\frac{d_{j,r}}{c_{i,r}}
\right) \right\}\right)
\eeqa
where 
$$\Mcal_j(\theta)=  
\int_{g_j(0)}^{g_j(S^{(j)}_{\max})} 
\log(\Phi_j(x)\mbox{e}^{\theta}
+ 1-\Phi_j(x))
\mbox{d}x +
\theta g_j(0)$$
and
$\Phi_j(x)  =   \P(g_j(S^{(j)})>x).$

\end{corollary}

{\it Proof:} For $\theta>0$,
\beqa
\lefteqn{
\log \E \exp\left(\theta \max_r \frac{\sum_j q_{j,r}}{c_{i,r}} \right) 
 } & & \\
& = & 
\log \E \exp\left(\theta  \max_r \sum_j \frac{Q_j d_{j,r}}{c_{i,r}} \right)\\
& \leq & 
\log \E \exp\left(\theta 
\sum_j Q_j \max_r\frac{d_{j,r}}{c_{i,r}} \right)\\
& = & 
\sum_j \log \E \exp\left(
\theta Q_j \max_r\frac{d_{j,r}}{c_{i,r}} 
\right),
\eeqa
where the last equality is by assumed mutual independence of the
$Q_j$, $j\in\Jcal$.
The proof then follows by the argument for the single-tier
theorem
\ifTR
(\ref{yes-blocking}) 
\cite{KesTas00} (also see Appendix A)
\else
(\ref{yes-blocking}) \cite{lambda-netcalc-arxiv,KesTas00} 
\fi
and the Chernoff bound. \qed

\subsection{An atomic service in terms of IT resources allocated}

Consider the special case of an atomic service tier in
terms of allocated resources (as in AWS Lambda). 
That is, suppose there are constants $\kappa_j$ such that 
\be\label{atomic-resource}
\forall j\in\Jcal,r\in\Rcal, ~~ d_{j,r} & = & \kappa_j d_{1,r}.
\ee
Regarding Corollary \ref{cor:mc} for this case, obviously
\beqa
\forall j, ~\max_r \frac{d_{j,r}}{c_{i,r}} & = & 
\kappa_j \max_r \frac{d_{1,r}}{c_{i,r}}.
\eeqa
Here, a tier-$j$ service instance would
consume $\kappa_j$ tokens upon invocation.

\subsection{Extensions to multiple physical servers}

To extend the case of multiple service tiers
to multiple physical servers $i$,
one can divide each tenant $n$'s demand envelope among
them. For example,
for nonnegative scalars $\alpha_{j(n),i}$ such that $\sum_i \alpha_{j(n),i}=1$,
take $g_{j(n),i}=\alpha_{j(n),i} g_{j(n)}$ so that 
$$g_{j(n)} = \sum_i g_{j(n),i}.$$
The weights $\alpha$ for
each tenant can then be chosen to balance load among servers $i$. Given that,
Corollary \ref{cor:mc} can be used for each server $i$.

Obviously, the above approach to admission control could be separately applied
to each tier in $\Jcal$ if
resources for different service tiers are statically partitioned
based on demand assessments.

Note that under (\ref{atomic-resource}), the price of 
type-$j$ Lambda service instances should be {\em more} than
$\kappa_j$ type-1 (atomic) Lambda service instances because the former needs
to be allocated on a single physical server.


\section{Future Work}\label{sec:fw}

For longer running Lambda functions, if 
Lambda servers are available, it may be more
resource efficient to invoke  
requests that violate their token-bucket profiles but
flag them \cite{2R3CM,1R3CM} as preemptible or  pausable.
Also, blocked in-profile and out-of-profile requests may be 
temporarily queued.
In future work, we will study the overhead of preemption
and the performance of
policies  to price and preempt out-of-profile invocations.

Nonlinear chance constraints can replace linear ``spatial" resource constraints
such as  (\ref{K-def}).
In future work, we will also consider how the above temporal approach to
overbooking can be combined with instance-placement 
approaches based on chance constraints.  


\ifTR
\section*{Acknowledgements}
This research was supported in part by NSF CNS 1717571 grants and a Cisco Systems URP gift. 
\else
\begin{acks}
This research was supported in part by NSF CNS 1717571 grants and a Cisco Systems URP gift. 
\end{acks}
\fi

\ifTR
\bibliographystyle{plain}
\else
\bibliographystyle{ACM-Reference-Format}
\fi
\bibliography{../../latex/scheduling,../../latex/kesidis-prior,../../latex/cloud0,../../latex/routing,../../latex/netcalc,../../latex/new}

\ifTR
\section{Appendix A: Loss system with arrivals satisfying burstiness curves}

In this Appendix, we reinterpret the statement of
Theorem 2 of \cite{KesTas00} and provide a modified proof.
Consider a bufferless system with $K\geq 1$ identical servers.
Let $T_i$ be the arrival time of job (service request)
$i$ and here let  $S_i$ be its service time.
 Consider a (increasing, concave and nonnegative) burstiness curve $g$ for arrivals,  
\ie 
$$\forall s\leq t,  ~~\sum_i {\bf 1}\{s < T_i\leq t\} \leq g(t-s).$$
Assume a maximum service time $S_{\max}$.

The number of busy servers (jobs in the system) at time $t$,
\beqa
Q(t) & = &  \sum_{i=-\infty}^\infty {\bf 1}\{T_i \leq t < T_i + S_i \}\\
 & = &  \sum_{i=-\infty}^\infty {\bf 1}\{t-S_i < T_i \leq t\}\\
& \leq & \sum_{i=-\infty}^\infty {\bf 1}\{t-S_{\max} < T_i \leq t\}\\
& \leq  & g(S_{\max}).
\eeqa
So, if $g(S_{\max})\leq K$, then the $K$-server system will never
block jobs \cite{Cruz95}.

\begin{theorem}\label{logmgf-bound}
\cite{KesTas00}
If
\be
\P(S=0) & =  & 0 \label{strictly-pos}
\ee
and the service times
$S_i$ are independent and identically distributed,
then in steady-state,
\beqa
\log \E \mbox{e}^{\theta Q} & \leq &  \theta g(0)+
\\ & & 
\int_{g(0)}^{g(S_{\max})} \log(\Phi(x)\mbox{e}^\theta + 1-\Phi(x))
\mbox{d}x  \\
& =: & \mathcal{M}(\theta)
\eeqa
where 
$\Phi(x)  =   \P(g(S)>x)$.
\end{theorem}

\begin{corollary}
If (\ref{strictly-pos}) and the $S_i$ are independent and identically
distributed, then in steady-state
the Chernoff bound is
\beqa
\P(Q>K) & \leq & \exp(-\sup_{\theta>0}\{ \theta K  - \mathcal{M}(\theta)\}).
\eeqa
\end{corollary}

\begin{corollary}\label{cor:Markov}
If (\ref{strictly-pos}) and  the $S_i$ are identically distributed,
then in steady-state
the Markov inequality is,
\beqa
\P(Q>K) & \leq & \frac{\E g(S)}{K}.
\eeqa
\end{corollary}

\noindent
{\it Remark:} For Corollary \ref{cor:Markov},
the $S_i$ are not necessarily mutually independent.


\noindent
{\it Proof of the Theorem:}
Define a partition $\{m_\ell\}_{\ell=0}^{L+1}$ of the range of $g$:
\beqa
m_0=g(0), ~m_\ell < m_{\ell+1} ~\forall \ell, ~m_{L+1}=g(S_{\max}).
\eeqa
Define the job indexes so that $T_{-1} \leq t < T_0$ and
\beqa
Q(t) & =  &  \sum_{i=-\infty}^{-1} {\bf 1}\{t-S_i < T_i \leq t\}.
\eeqa
Thus, 
\beqa
Q(t) & =  &  \sum_{i=-\infty}^{-1} \sum_{\ell=0}^L
{\bf 1}\{t-S_i < T_i \leq t\}
\\ & & ~~
\cdot
{\bf 1}\{g^{-1}(m_\ell)< S_i \leq g^{-1}(m_{\ell+1})\}\\
& \leq & \sum_{\ell=0}^L \sum_{i=-\infty}^{-1} 
{\bf 1}\{t-g^{-1}(m_{\ell+1}) < T_i \leq t\}
\\ & & ~~
\cdot
{\bf 1}\{g^{-1}(m_\ell)< S_i \leq g^{-1}(m_{\ell+1})\}\\
& \leq & \sum_{\ell=0}^L \sum_{i=-m_{\ell+1}}^{-1} 
{\bf 1}\{g^{-1}(m_\ell)< S_i \leq g^{-1}(m_{\ell+1})\}
\eeqa
where the inequalities are by the burstiness constraint $g$ on 
$\{T_i\}$.


Switching the order of summation again gives,
\beqa
Q(t) &\leq  & \sum_{i=-m_1}^{-1} 
\sum_{\ell=0}^L 
{\bf 1}\{g^{-1}(m_\ell)< S_i \leq g^{-1}(m_{\ell+1})\}\\
& & ~~+ \sum_{i=-m_2}^{-m_1-1} 
\sum_{\ell=1}^L 
{\bf 1}\{g^{-1}(m_\ell)< S_i \leq g^{-1}(m_{\ell+1})\}\\
& & ~~+ ...  + \sum_{i=-m_{L+1}}^{-m_L-1} 
{\bf 1}\{g^{-1}(m_L)< S_i \leq g^{-1}(m_{L+1})\}\\
& = & \sum_{i=-m_1}^{-1} {\bf 1}\{g^{-1}(m_0)< S_i\}\\
& & ~~+ \sum_{i=-m_2}^{-m_1-1} {\bf 1}\{g^{-1}(m_1)< S_i\}\\
& & ~~ + ...  + \sum_{i=-m_{L+1}}^{-m_L-1} 
{\bf 1}\{g^{-1}(m_L)< S_i \}.
\eeqa

Taking expectation now and letting the partition 
$\{m_{\ell}\}_{\ell=0}^{L+1}$ become infinitely fine as $L\rightarrow\infty$
leads to $\E Q\leq \E\int_{g(0)}^{g(S_{\max})}
\P(g(S)>x)\mbox{d}x = \E g(S)$ and Corollary \ref{cor:Markov}.

Continuing from the previous display:
Since the $S_i$ are identically distributed $\sim S$, 
\beqa
\forall i, ~~\E \exp(\theta {\bf 1}\{g^{-1}(m_\ell)< S_i \}) & =&  \Phi(m_\ell)\mbox{e}^\theta+1-
\Phi(m_\ell).
\eeqa
Since $S_i$ are independent and $\Phi(m_0)=1$
(the latter because $\P(S=0)=0$),
\beqa
\E \mbox{e}^{Q (t)} & \leq & 
\mbox{e}^{\theta  m_0}
\prod_{\ell=0}^L 
(\Phi(m_\ell) \mbox{e}^\theta + 1- \Phi(m_\ell))^{m_{\ell+1}-m_\ell} 
\eeqa
Thus,
\beqa
\lefteqn{\log \E \mbox{e}^{Q (t)} ~ \leq ~ \theta g(0)} & & \\
& + &
 \sum_{\ell=0}^L (m_{\ell+1}-m_\ell)
\log (\Phi(m_\ell) \mbox{e}^\theta + 1-\Phi(m_\ell))
\eeqa
So, as $L\rightarrow \infty$ and the partition $\{m_\ell\}$ of the
range of $g$ becomes infinitely fine,
this bound converges to the integral,
\beqa
\theta g(0)+
\int_{g(0)}^{g(S_{\max})} \log(\Phi(x)\mbox{e}^\theta + 1-\Phi(x))
\mbox{d}x  
~~~~\qed
\eeqa

\fi

\end{document}